# Ultra-low noise dual-frequency VECSEL at telecom wavelength using fully correlated pumping

Hui Liu,[1] Gregory Gredat,[1] Syamsundar De,[1] Ihsan Fsaifes,[1] Aliou Ly,[1] Rémy Vatré,[1] Ghaya Baili,[2] Sophie Bouchoule,[3] Fabienne Goldfarb,[1] and Fabien Bretenaker[1,*]

[1]*Laboratoire Aimé Cotton, CNRS – Université Paris-Sud – Ecole Normale Supérieure Paris-Saclay, Université Paris-Saclay, Orsay, France*
[2]*Thales Research & Technology, Palaiseau, France*
[3] *Centre de Nanosciences et Nanotechnologie (C2N), CNRS - Université Paris-Sud – Université Paris-Saclay, Marcoussis, France*
*Corresponding author: fabien.bretenaker@u-psud.fr

**An ultra-low intensity and beatnote phase noise dual-frequency vertical-external-cavity surface-emitting laser is built at telecom wavelength. The pump laser is realized by polarization combining two single-mode fibered laser diodes in a single-mode fiber, leading to a 100 % in-phase correlation of the pump noises for the two modes. The relative intensity noise is lower than -140 dB/Hz, and the beatnote phase noise is suppressed by 30 dB, getting close to the spontaneous emission limit. The role of the imperfect cancellation of the thermal effect resulting from unbalanced pumping of the two modes in the residual phase noise is evidenced.**

Vertical-external-cavity surface-emitting lasers (VECSELs) constitute a flourishing research topic, because these lasers provide lots of advantages such as a wide spectral range, high beam quality, high output power, and low noise properties [1, 2]. Thanks to their external cavity, such lasers exhibit strong versatility in their designs and applications. One example is the dual-frequency VECSEL (DF-VECSEL) [3], which is an attractive way to produce an optically carried single-side-band radio frequency (RF) signal with a broad tunability. Thanks to the rejection of noises that are common to the two modes, the generated beatnote exhibits a remarkable spectral purity thanks to the laser class-A dynamics [4, 5]. This makes the phase locking of the generated beatnote to an external reference source much more simple than in the case of two independent lasers. DF-VECSELs for the generation of optically carried RF source have been intensively studied for microwave photonics applications [6] and could find applications in optical telecommunications, radars, and lidars [7]. Further, the emission of two cross-polarized laser frequencies with an adjustable frequency difference is advantageous for application to the field of miniaturized atomic clocks based on coherent population trapping resonances [8].

Some efforts have been made to develop DF-VECSELs at several wavelengths, including 1 µm [3], 1.55 µm for microwave photonics applications [9], and 852 nm for cesium atomic clocks [8]. Furthermore, the noise mechanisms in DF-VECSEL have been analyzed [10-12], showing that the intensity noises of the two modes and the phase noise of the beatnote between the two orthogonally polarized modes mainly come from the pump noise. All these previously reported DF-VECSELs used multimode fiber coupled pump lasers in order to reach a high pump power. This multimode guiding of the pump light is detrimental to the pump noise, which is deteriorated by the fluctuations of the speckle pattern resulting from the interferences between the large number of modes in the fiber. Moreover, this speckle decreases the correlations between the pump noises seen by the two modes of the DF-VECSEL, which are slightly spatially separated in the active medium to reduce gain competition [12]. These two effects lead to a strong increase of the beatnote phase noise. This is particularly undesirable because the bandwidth of these noises exceeds several tens of MHz and is thus very difficult to reduce with an optical phase lock loop.

The theory developed in Ref. 12 predicts that the intensity noises and the beatnote phase noise induced by the pump noise do not depend only on the pump intensity noise level. More precisely, it predicts that the beatnote phase noise can be drastically reduced by increasing the correlations between the noises of the pumping intensities that feed of the two lasing modes, provided the phase of this correlation is 0 (so-called in-phase correlation). In order to increase the correlations of the intensity fluctuations of two transversally spatially separated regions of the same beam, it seems natural to try to use a single spatial mode, which exhibits maximum transverse coherence.

However, the main problem with single-mode fibered pump lasers is that because of their relatively small numerical aperture, they cannot deliver as much power as multimode-fibered lasers. Thus, we combine two 950 mW pump lasers emitting at 976 nm fibered with polarization maintaining fibers, using a fibered polarization combiner, as shown in Fig. 1(a). In these conditions, about 1.8 W of laser power is available in the output single-mode polarization maintaining fiber. Then, the pump laser is focused onto the semiconductor structure by two lenses forming a telescope.

The schematic of our DF-VECSEL is depicted in Fig. 1(b). The laser operating at 1.56 µm is based on a 1/2-VECSEL structure [5]. This structure consists of an InP-based ~4λ-thick active region, including eight strained InGaAlAs quantum wells distributed among 3 optical standing wave antinode positions with 4-2-2 distribution. A 17-pair $GaAs/Al_{0.97}Ga_{0.03}As$ semiconductor distributed Bragg reflector is integrated to the active region using a metamorphic regrowth, completed with a gold layer in order to reach a reflectivity larger than 99.9% at 1.55 µm. The overall gain structure is bonded on a chemical vapor deposition (CVD) polycrystalline diamond substrate [13]. A quarter-wavelength antireflection (AR) layer at the laser wavelength is finally deposited on the sample surface. This coating leads to a 25 % reflectivity at the pump wavelength. The temperature of the 1/2-VECSEL structure is controlled at 20°C using a Peltier thermoelectric cooler.

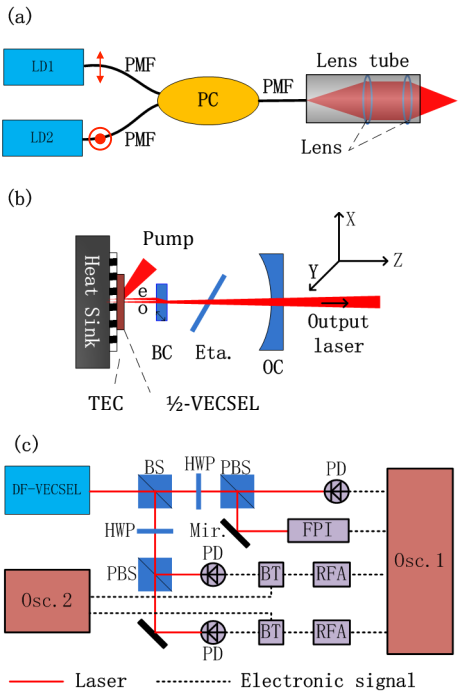

Fig. 1. Experimental setup. (a) The two pump laser diodes LD1 and LD2 are combined by the fibered polarization combiner PC. PMF: polarization maintaining fiber. (b) VECSEL structure. OC: output coupler; BC: birefringent crystal; TEC: thermoelectric controller. (c) Laser analysis setup. PBS: polarization beam splitter; FPI: Fabry-Perot interferometer; HWP: half-wave plate; PD: photodiode; BT: bias tee; RFA: low-noise radio frequency amplifier.

The laser cavity is closed by a concave output coupler with a reflectivity of 99.4 % and a radius of curvature equal to 5 cm. The length of the cavity is 49.0 mm, leading to a mode waist radius equal to 59 µm on the chip. The cavity contains a 0.5-mm-thick anti-reflection-coated $YVO_4$ birefringent crystal cut at 45° of its optical axis. This crystal creates a 50 µm transverse walk-off between the ordinary y-polarized and extraordinary x-polarized modes in the gain structure. This partial spatial separation reduces gain competition between the two modes, allowing them to oscillate simultaneously inside the laser. Finally, an uncoated YAG etalon with a thickness of 100 µm selects a single longitudinal mode for each polarization. As shown in Fig. 1(c), the laser output is divided into several beams to simultaneously perform the different measurements. One of the beams is used to monitor the laser spectrum using a Fabry-Perot interferometer. The intensities of the two cross-polarized modes and the beatnote signal between them can be simultaneously detected and recorded with a fast deep memory oscilloscope.

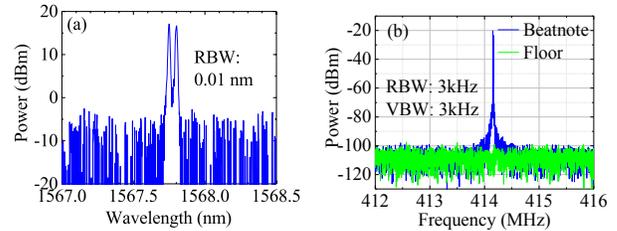

Fig. 2. (a) DF-VECSEL spectrum measured by an optical spectrum analyzer (resolution: 0.01nm). (b) Beatnote spectrum measured with an electrical spectrum analyzer.

A typical optical spectrum, recorded by an optical spectrum analyzer with a resolution of 0.01 nm, is displayed in Fig. 2(a). The presence of two peaks evidences the dual-frequency oscillations. The frequency difference between the two modes can be tuned between 0 and 3 GHz by slightly rotating the birefringent crystal. Fig. 2(b) shows an example of beatnote signal obtained when this frequency difference is tuned down to 414 MHz. A single peak with a width limited by the resolution bandwidth of the spectrum analyzer illustrates the single-mode behavior of each mode. Also, it is worth noticing that this beatnote has a very small and narrow-bandwidth noise pedestal compared to preceding results obtained with spatially multimode pumping [9].

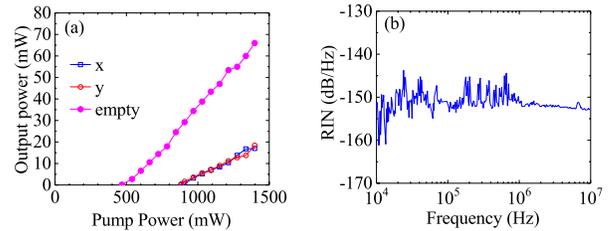

Fig. 3. (a) Evolution of the VECSEL output power versus incident pump power. Full points: empty cavity. Open symbols: output powers of the x- and y-polarized modes in dual-frequency configuration. (b) Pump laser RIN spectrum.

The dependence of the DF-VECSEL output power on the pump power is shown in Fig. 3 (a). One can see that with a pump power equal to 1.4 W, the output power is reduced from 65 mW to 15 mW when we insert the birefringent crystal and the etalon inside the cavity. The relatively low efficiency of this laser, even with an empty cavity, is due to the small gain of such structures at 1.5 µm and to the fact that the output coupler transmission is not optimized for empty cavity operation. Although a 15 mW output power is sufficient for some applications, larger powers could be obtained by reducing the reflection coefficient of the chip at the pump wavelength and using a resonant microcavity 1/2-VECSEL architecture [14], provided such a microcavity does not increase the competition between the modes.

To analyze the laser noises, we use the experimental setup sketched in Fig. 1 (c). All noises are analyzed in the 10 kHz to 10 MHz frequency range, in which we suppose that all random processes are stationary. The spectra of the relative intensity noises (RINs) of the two modes are shown by the scattering plots in Fig. 4 (a). As one can see, the RINs of both modes lie below the level of -140 dB/Hz, indicating a large intensity noise reduction compared with results obtained with spatially multimode pumps [9], which were typically located between -120 dB/Hz to -130 dB/Hz.

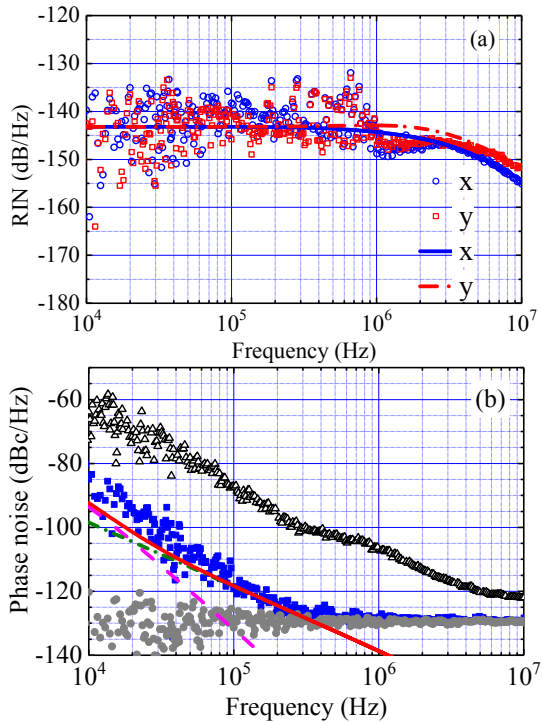

Fig. 4. Noise spectra. (a) RIN spectra of the two laser modes. The symbols correspond to the measurements while the lines are obtained from theory. (b) PSD of the beatnote phase noise. Filled squares (resp. open triangles): data obtained with spatially monomode (resp. multimode) pumping. Dashed line: calculated thermal component of the phase noise. Dotted-dashed line: contribution predicted from Schawlow-Townes linewidth. Full line: sum of the two contributions. Closed circles: measurement floor. The parameters used in the modeling are: excitations ratios of the two modes $r_x = r_y = 1.57$; photon lifetimes for the two modes $\tau_x = 20$ ns, $\tau_y = 15$ ns; carrier lifetime $\tau = 1$ ns, pump RIN $RIN_P = -152$ dB/Hz, pump powers $P_{P,x} = 0.5$ W, $P_{P,y} = 0.67$ W for the two modes, thermal resistance $R_T = 24$ K/W, relative thermal variation of the cavity length $\Gamma_T = 2.0 \times 10^{-8}$ K$^{-1}$, thermal response time $\tau_T = 23$ μs, phase-amplitude coupling factor $\alpha = 7$, steady-state number of intracavity photons for the two modes $F_{0x} = F_{0y} = 5.2 \times 10^9$.

These spectra are in good agreement with the theoretical ones (see the lines in Fig. 4 (a)). These theoretical spectra were computed using the expressions derived in Ref. 12 with a pump RIN level equal to $RIN_P(f) = -152$ dB/Hz at all noise frequencies $f$, obtained from the measurement shown in Fig. 3(b). This modeling also takes into account the correlation between the pump noises seen by the two laser modes. Indeed, because of their partial spatial separation, the pump intensity fluctuations $\delta P_{P,x}(t)$ and $\delta P_{P,y}(t)$ seen by the two modes are usually different, leading to a partial correlation of their Fourier transforms $\delta \tilde{P}_{P,x}(f)$ and $\delta \tilde{P}_{P,y}(f)$. This correlation is a complex quantity that has a modulus $\eta(f)$ and a phase $\Psi(f)$ defined as:

$$\eta(f)e^{i\Psi(f)} = \frac{\langle \delta \tilde{P}_{P,x}^*(f) \delta \tilde{P}_{P,y}(f) \rangle}{\sqrt{\langle |\delta \tilde{P}_{P,x}(f)|^2 \rangle \langle |\delta \tilde{P}_{P,y}(f)|^2 \rangle}}. \quad (1)$$

Contrary to preceding experiments that used multimode-fibered pump lasers, the use here of single-mode fiber to guide the pump light leads to a correlation amplitude between the two pump noises equal to $\eta = 1$ with a correlation phase $\Psi = 0$ at all considered frequencies $f$. The photon lifetimes of the two $x$- and $y$-polarized modes are taken to be $\tau_x = 15$ ns and $\tau_y = 20$ ns. Indeed, the two modes experience slightly different losses because of the presence of a tilted etalon and an imperfectly AR-coated birefringent crystal inside the cavity. As shown by the measurements of Fig. 3(a), we are able in the experiment to equalize the powers of the two orthogonally polarized modes by carefully adjusting the position of the pump spot on the semiconductor structure. This permits to compensate the loss difference between the two modes by a difference in pumping, leading to identical excitation ratios. For the modeling, we take these excitation ratios equal to $r_x = r_y = 1.57$. The carrier lifetime is $\tau = 1$ ns. The coupling constant between the two modes is equal to $C = 0.39$. The cut-off frequency at about 2 MHz is due to the filtering effect of the photon lifetime in the cavity, thus corroborating the laser class-A dynamics.

The corresponding beatnote phase noise spectrum is shown by the full squares in Fig. 4(b). For comparison, we reproduce in the same plot the phase noise of the DF-VECSEL pumped with a spatially multimode pump (see the open triangles). This spatially multimode pump beam exhibited a RIN equal to -145 dB/Hz in the considered frequency range. This relatively strong intensity noise was associated with a poor correlation amplitude $\eta = 0.6$ between the pump noises undergone by the two modes in the considered frequency range. As can be seen, spatially monomode pumping leads to a reduction of more than 20 dB of the phase noise between 10 kHz and several MHz. In particular, the bandwidth of the remaining phase noise, which is still measurable above the −130 dB/Hz measurement floor, is reduced from more than 10 MHz to less than 200 kHz, a bandwidth that can easily be managed by inserting the laser in an optical phase locked loop.

To compare the phase noise measurements with theory, we insert the pump noise measured in Fig. 3(b) into the model reported in Ref. 12. In this model, the phase noise of the beatnote is due to two different sources: i) the thermal effect in which the pump noise induces temperature fluctuations of the chip that are transferred to the phases of the two modes through the thermal dependence of the refractive index of the semiconductor structure and ii) the phase amplitude coupling mechanism that transfers the intensity noises to the phase noises of the two modes. In particular, the thermal component of the power spectral density of the beatnote phase noise is given, in the case where the two pumps are perfectly correlated ($\eta = 1$), by [12]:

$$S_T(f) = \left(\frac{\omega_0 \Gamma_T R_T}{2\pi f}\right)^2 \frac{\text{RIN}_P(f)}{1+4\pi^2 f^2 \tau_T^2}(P_{P,x} - P_{P,y})^2, \quad (2)$$

where $P_{px}$ and $P_{py}$ are the pump powers for the two modes, $\omega_0$ is the laser angular frequency, $R_T$ and $\tau_T$ are the thermal resistance and response time of the semiconductor structure for the considered pump mode diameter. $R_T$ has been measured. $\tau_T$ is evaluated at 23 µs thanks to the knowledge of the pump beam radius on the structure ($w_P = 75$ µm) and the thermal diffusion coefficient ($D_T = 3.7 \times 10^{-5}$ m²/s). $\Gamma_T$ is the relative thermal variation of the cavity optical length when the chip temperature varies, given by $\Gamma_T = (L_{SC}/L_{ext})d\bar{n}/dT$ where $L_{SC}$ is the length of the active region, $L_{ext}$ the length of the external cavity, and $d\bar{n}/dT$ the thermo-optic coefficient of the active region, obtained by taking the average of the thermo-optic coefficients of the materials of the active region weighted by their thicknesses. In Eq. (2), we suppose that the pump fluctuations injected into the two modes are fully in-phase correlated ($\eta = 1$). We can thus see that the only remaining component in the thermal phase noise is due to the residual unbalance between the two pump powers seen by the two modes. Equation (2) leads to the dashed line of Fig. 4(b). The good agreement with the measurements confirms the fact that the phase/intensity coupling effect is negligible in the case of single-mode pumping, thanks to the perfect correlation of the two pumps ($\eta = 1$). From Eq. (2), one could expect a better control of the pump balance between the two modes to allow a strong further reduction of the phase noise. However, one can see from the dotted-dashed line of Fig. 4(b) that the noise level that we have reached is no longer far from the phase diffusion induced by the Schawlow-Townes linewidth of the two modes, which is given (in rad²/Hz) by $S_{ST}(f) = S_{ST,x}(f) + S_{ST,y}(f)$ [4,15] with

$$S_{ST,i}(f) = \frac{1}{8\pi^2 f^2 F_{0i}\tau_i}\left\{1 + \frac{\alpha^2(r_i-1)\left[(r_i-1)+4\pi^2 f^2 r_i \tau_i^2\right]}{(r_i-1-4\pi^2 f^2 \tau \tau_i)^2 + 4\pi^2 f^2 \tau_i^2 r_i^2}\right\}, (3)$$

for $i=x, y$ and where the values of the parameters are listed in the caption of Fig. 4, and where we have treated the two modes as two independent lasers. Figure 4(b) shows that the noise induced by the pump is now very close to spontaneous emission noise that will be the next limit to overcome in such experiments.

In conclusion, an ultra-low noise DF-VECSEL has been demonstrated. Low noise is achieved by pumping this laser with a low noise pump exhibiting 100 % in-phase correlation between spatially separated regions of the pump beam. This was made possible by polarization combining two single-mode-fibered pump lasers into the same single-mode fiber. With respect to the case where the pump laser is coupled to a multimode fiber, more than 10 dB improvement is observed on the laser RIN and more than 20 dB on the phase noise of the beatnote. The residual phase noise of the beatnote comes from the unbalanced pumping of the two lasing modes and from spontaneous emission. Further noise reduction could thus be obtained with a better scheme allowing a more precise balancing of the losses of the two modes. However, in the present experiment already, the frequency range in which the residual phase noise is larger than– 130 dBc/Hz, is reduced from more than 10 MHz to less than 200 kHz, opening the way to a strong noise reduction by implementing an optical phase locked loop. Moreover, all those results are perfectly reproduced by our model for the noises of DF-VECSELs.

**Funding.** Agence Nationale de la Recherche (ANR) (ANR-15-CE24-0010-04, ANR-15-ASMA-0007-04); Direction Générale de l'Armement (DGA).

**Acknowledgment**. This work is performed in the framework of the joint research laboratory between Thales Research & Technology and Laboratoire Aimé-Cotton. C2N laboratory belongs to RENATECH, the French national network of large facilities for micronanotechnology. We acknowledge the help of Dominique Papillon for splicing the couplers and the help of Jean-Louis Loureau in purchasing the pump lasers.